# Deep Learning-based Automated Aortic Area and Distensibility Assessment: The Multi-Ethnic Study of Atherosclerosis (MESA)

Vivek P. Jani, MS[1], Nadjia Kachenoura, PhD[2], Alban Redheuil, MD, PhD[2], Gisela Teixido-Tura, MD, PhD[3], Kevin Bouaou, MS[2], Emilie Bollache[2], PhD, Elie Mousseaux, MD[4], Alain De Cesare, PhD[2], Shelby Kutty, MD, PhD[1], Colin O. Wu, PhD[5], David A. Bluemke, MD[6], Joao A. C. Lima, MD[1], Bharath Ambale-Venkatesh, PhD[1]

[1]Johns Hopkins University,
Baltimore, MD

[2]Sorbonne Université, Laboratoire d'Imagerie Biomédicale, INSERM, CNRS,
Paris, France

[3]Hospital Vall d'Hebron,
Barcelona, Spain

[4]Université de Paris, Hôpital Européen Georges Pompidou, APHP, INSERM PARCC, Paris, France

[5]Office of Biostatistics Research, Division of Intramural Research
National Heart, Lung and Blood Institute
National Institutes of Health
Bethesda, MD

[6]University of Wisconsin School of Medicine and Public Health
Madison, WI

Correspondence:
Bharath Ambale-Venkatesh, PhD
Assistant Professor, Department of Radiology
Johns Hopkins University
600 North Wolfe St
Baltimore, MD 21287
Ph: 443-287-3248 | bamabale1@jhmi.edu

**Acknowledgements.** The authors thank the other investigators, the staff, and the participants of the MESA study for their valuable contributions. A full list of participating MESA investigators and institutions can be found at http://www.mesa-nhlbi.org.

**Grant Support.** This research was supported by contracts HHSN268201500003I, N01-HC-95159, N01-HC-95160, N01-HC-95161, N01-HC-95162, N01-HC-95163, N01-HC-95164, N01-HC-95165, N01-HC-95166, N01-HC-95167, N01-HC-95168 and N01-HC-95169 from the National Heart, Lung, and Blood Institute, and by grants UL1-TR-000040, UL1-TR-001079, and UL1-TR-001420 from the National Center for Advancing Translational Sciences (NCATS).

**ABSTRACT**

**Background:** This study applies convolutional neural network (CNN)-based automatic segmentation and distensibility measurement of the ascending and descending aorta from 2D phase-contrast cine magnetic resonance imaging (PC-cine MRI) within the large MESA cohort with subsequent assessment on an external cohort of thoracic aortic aneurysm (TAA) patients.

**Methods:** 2D PC-cine MRI images and corresponding analysis of the ascending and descending aorta at the pulmonary artery bifurcation from the MESA study were included. Train, validation, and internal test sets consisted of 1123 studies (24282 images), 374 studies (8067 images), and 375 studies (8069 images), respectively. An external test set of TAAs was included, which consisted of 37 studies (3224 images). A U-Net based CNN was constructed, and performance was evaluated utilizing dice coefficient (for segmentation) and concordance correlation coefficients (CCC) of aortic geometric parameters by comparing to manual segmentation and parameter estimation.

**Results:** Dice coefficients for aorta segmentation were 97.6% (CI: 97.5%-97.6%) and 93.6% (84.6%-96.7%) on the internal and external test of TAAs, respectively. CCC for comparison of manual and CNN based maximum and minimum and ascending aortic areas were 0.97 and 0.95, respectively, on the internal test set and 0.997 and 0.995, respectively, for the external test. CCCs for maximum and minimum descending aortic areas were 0.96 and 0.98, respectively, on the internal test set and 0.93 and 0.93, respectively, on the external test set. The absolute differences between manual and CNN-based ascending and descending aortic distensibility measures were 0.02±9.7 and 20±10 $10^{-4}mmHg^{-1}$, respectively, on the internal test set, and 0.44±20 and 20±10 $10^{-4}mmHg^{-1}$, respectively, on the external test set.

**Conclusion:** We successfully developed and validated a U-Net based ascending and descending aortic segmentation and distensibility quantification model in a large multi-ethnic database and in an external cohort of TAA patients.

**INTRODUCTION**

Increased arterial stiffness is associated with aging and incident cardiovascular disease, namely stroke, ischemic heart disease, and heart failure (1–4). A stiffened aorta provides less systolic cushioning resulting in increased systolic blood pressure and inducing left ventricular hypertrophy (2, 5). Population and community-based studies in patients have demonstrated that aortic distensibility is an independent predictor of fatal and non-fatal cardiovascular events (3, 5, 6). Furthermore, direct measurements of central aortic stiffness by means of aortic distensibility have been shown to be an early marker of subclinical vascular alterations (5, 7). Cardiovascular magnetic resonance (CMR) has the unique ability to simultaneously assess aortic stiffness and ventricular function (1). Recently, convolution neural networks (CNNs) have demonstrated remarkable performance for classification, segmentation, and prediction tasks in radiology and CMR related tasks (8–13). However, deep learning applications for aortic segmentation are limited by lack of models trained on large cohorts of data required for robust, generalizable development and widespread use.

Previously, in MESA, the Art-Fun software has been used for aortic analysis, which has been validated on a pulsatile phantom and in several human studies; however, manual intervention is still required (1). This study details application of U-Net, a CNN-based arhictecture, for automatic segmentation of the ascending and descending aorta from 2D PC-cine MRI for automatic aortic analysis, including completely automated quantification of aortic cross-sectional areas and aortic distensibility. We subsequently tested the efficacy of our model on a separate cohort of patients with thoracic aortic aneurysms (TAA) to demonstrate application to clinically relevant pathologies.

**METHODS**

**MESA Population Characteristics.** The MESA (Multi-Ethnic Study of Atherosclerosis) is a prospective cohort study that evaluates both risk factors and mechanisms underlying cardiovascular disease progression and development among asymptomatic individuals (14). A total of 6418 individuals without CVD aged 45 to 84 years, and identified as white, black, Hispanic, or Chinese were recruited between 2000 and 2002 from 6 US field centers (Wake Forest University (WFU), Winston-Salem, NC; Columbia University (COL), New York, NY; Johns Hopkins University (JHU), Baltimore, MD; University of Minnesota, Twin Cities (UMN), Minneapolis, MN; Northwestern University (NWU), Chicago, IL; University of California, Los Angeles (UCLA), Los Angeles, CA). Characteristics of the population with suitable aortic MRI were 48% men, 34% white, 14% Chinese America, 30% African American, and 22% Hispanic, with mean age of 64±10 years. Of 5005 participants with CMR imaging, 1872 had 2D Phase Contrast cine MRI images of the aorta at the main pulmonary artery bifurcation with corresponding analysis and segmentations. No studies were excluded on the basis of image quality. All participants gave informed consent for the study protocol, which was approved by the institutional review boards of all MESA field centers and the CMR reading center.

**MESA MRI Imaging.** MESA MRI images were acquired with 1.5T whole-body scanners and gradient echo phase-contrast cine MRI with electrocardiographic gating for aortic flow and lumen evaluation (5). Specifically, the following scanners were used: GE CV/i or LX, Waukesha, WI and Siemens Symphony/Sonata, Erlangen, Germany. Ascending and descending aorta images were obtained in the transverse plane perpendicular to the aortic centerline at the level of main pulmonary artery bifurcation. Imaging parameters were as follows: repetition time– 10 ms; echo time – 1.9 ms; flip angle – 20 degree; field of view – 340 mm; slice thickness – 8 mm; matrix – 256x256; number of images – 20 for 1 cardiac cycle; encoding velocity – 150 cm/s; bandwidth – 245 Hz/pixel.

**External Test Set Characteristics:** A total of 37 individuals (3224 images) from a cohort of TAAs external to MESA were included. Characteristics of the population with suitable aortic MRI are shown in **Table 1.** Central pressures that were recorded simultaneously to MRI acquisitions using Sphygmocor and are shown in **Table 1**. A total of 19 out of the 37 individuals had a bicuspid aortic valve (BAV). Three of these individuals had a dilated ascending aorta but did not meet the aneurysmal threshold. All remaining 18 patients with a tricuspid aortic valve had TAA. No patients in this cohort had an aortic dissection.

**External Test Set MRI Imaging:** TAA MRI images were acquired with 3T whole-body scanners and gradient echo phase-contrast cine MRI with electrocardiographic gating for aortic flow and lumen evaluation. Ascending and descending aorta images were obtained in the transverse plane perpendicular to the aortic centerline at the level of main pulmonary artery bifurcation. Imaging parameters were as follows: repetition time– 5.6 ms; echo time – 3.5 ms; flip angle – 20-25°; field of view – 210-460 x 105-230 mm²; slice thickness – 5-8 mm; matrix – 256x128; number of images – 50-121 for 1 cardiac cycle depending on patient heart rate; encoding velocity – 160-400 according to the suspected presence of aortic valve stenosis; bandwidth – 326-561 Hz/pixel.

**Deep Learning Model and Training.** Automated segmentation utilized the U-Net CNN architecture, described elsewhere (15). Binary and multi-class segmentation models, in which the ascending and descending aorta were identified as separate classes, were trained with and without training data augmentation, which included random zoom, rotation, crop, and horizontal/vertical shifts. 2D PC-Cine MRI magnitude images from the MESA cohort were randomly assigned utilizing Python to non-overlapping training (60%), validation (20%), and internal test (20%) groups, as detailed in **Figure 1**. Train, validation, and *internal test sets* consisted of 1123 studies (24282 images), 374 studies (8067 images), and 375 studies (8069 images), respectively. We further included an *external test set* of TAA patients, which consisted of 37 studies (3224 images). Each image from a given time series was an independent data set, and time series information was not considered by the CNN. Images from the same participant and time series were assigned to the same train, validation, or internal/external test group. Segmentation maps for training were generated utilizing ArtFun and manually adjusted as previously detailed (5, 7, 16). Both images and segmentation maps were resized to 256x256 with zero-padding, if necessary.

An un-weighted categorical cross entropy loss function with Adam optimization was utilized for training (17). Network hyperparameters were as follows: learning rate – 1 x 10⁻

[5]; batch size – 64; epochs – 75. Hyperparameters were tuned utilizing empirical methods by maximizing the validation set accuracy, assessed from the dice coefficient and loss function. Learning rate and epochs parameters were updated in one-log increments, while the batch size parameter was updated by multiples of two. The model was built and trained in Python utilizing Keras with a TensorFlow backend. Model training and evaluation were performed on a server workstation with 12 CPU cores, 32 GB RAM, and two NVIDIA graphics processing units (GPUs) each with 16 GB of video memory (NVIDIA [Santa Clara, California, USA] Tesla P100).

**Model Evaluation.** Automated segmentation maps in the test set were evaluated relative to manual ground truth segmentation maps with the Dice coefficient (12, 18, 19). Briefly, the Dice coefficient assesses the overlap between two different segmentation maps, considering only non-zero background pixels. For visualization, contours were obtained from segmentation maps utilizing the OpenCV image processing library in Python.

**Quantification of Aortic Distensibility**. Aortic geometric parameters evaluated in this study include ascending and descending aortic area at systole and diastole, and ascending and descending aortic strain. Aortic strain was calculated as the difference of maximum and minimum aortic area divided by the minimum aortic area. Consequently, manually derived and deep learning derived segmentations resulted in different phases of systole and diastole, though these were not significant. Distensibility was quantified as strain over pulse pressure. Quantitative aortic parameters were obtained from the ArtFun software by providing as input manual and deep learning derived segmentation maps of the modulus images which were then superimposed to velocity images acquired from PC-cine MRI (5, 16). Aortic geometric parameters were quantified for two sets of contours for each participant, namely (1) contours defined by the user (manual) and (2) deep learning derived contours (deep learning).

**Statistical Analysis.** Continuous variables are expressed as mean ± SD, unless otherwise specified. Categorical variables are expressed as percentages. 95% confidence intervals for dice coefficients were generated assuming a normal distribution.

Agreement between geometric parameters calculated from deep learning and manual segmentation maps was assessed utilizing Bland-Altman analysis and the concordance-correlation coefficient (CCC) (20). All statistical analyses were performed using Stata, version 15.0 (Stata Corp LP, College Station, TX).

**RESULTS**

**Participant Characteristics**. Demographic and relevant clinical parameters for both MESA participants and TAA patients with suitable 2D Phase Contrast cine MRI of the aorta are presented in **Table 1**. Train, validation, and internal (MESA) test group characteristics were observed to be similar to population characteristics. No significant differences were observed in all other relevant clinical parameters between the train, validation, and internal (MESA) test groups.

**Model Training and Internal Evaluation**. Segmentation with and without augmentation of the training data set, was evaluated on an internal test set, randomly selected from MESA. Dice coefficients for these models are shown in **Supplemental Table 1** and **Figure 2** for both models. These data demonstrate that all models are in good agreement with manually segmented contours (Dice coefficients>90% for all). Of the two models, the segmentation model trained on non-augmented training data was utilized for further evaluation, as this model had the best performance and was successfully able to discern the ascending and descending aorta. Representative contours generated from this model are shown in **Figure 3**. To demonstrate generalizability, the model was evaluated across different sites in MESA, shown in **Supplemental Table 2.** These data demonstrated that the site/scanner of image acquisition did not appreciably affect model performance.

To further evaluate model accuracy, quantitative aortic parameters, namely time-resolved ascending and descending aorta area, maximum and minimum ascending aorta area, maximum and minimum descending aorta area, and ascending aorta and descending aorta distensibility, were evaluated for the internal MESA derived test. Bland

Altman plots for ascending and descending aortic areas and distensibility are shown in **Figures 4** and **5,** respectively. The mean difference and concordance correlation coefficients (CCC) between aortic parameters derived from manual and deep learning segmentations are shown in **Table 2**. These data demonstrate that model had excellent performance for quantification of aortic areas (CCC>90%), and fair-to-good performance for distensibility assessment.

**External Model Evaluation.** Following evaluation on the internal test set derived from MESA, multi-class models were evaluated on a test set of thoracic aortic aneurysm patients external to MESA. Model performance, in terms of dice coefficient is shown in **Figure 2** with representative contours shown in **Figure 3**. As with the internal data set, quantitative aortic parameters were evaluated for this external data set, with Bland Altman plots shown in **Figures 4** and **5** and CCC shown in **Table 3**. Similar to what was observed for the internal data set, the model had excellent performance for quantification of aortic areas (CCC>90%), and fair-to-good performance for quantification of distensibility. In particular, descending aortic distensibility had a lower CCC compared with that from the internal MESA derived data set.

**DISCUSSION**

This study evaluates the application of U-Net for automatic segmentation of the ascending and descending aorta and automatic quantification of aortic area and distensibility. Importantly, all deep learning models evaluated in this study were trained on the MESA data set, which is a large data set with a diverse population, and evaluated on an external cohort of thoracic aortic aneurysm patients, which are known to have appreciably different aortic geometry. Segmentation maps from U-Net were found to be in close agreement with manual segmentation maps as shown by the high dice coefficients for both test sets. Furthermore, aortic areas from deep learning segmentations were found to be in excellent agreement with those from manual segmentations. However, the model only had moderate correlation for quantification of aortic distensibility when compared to manual segmentations.

Our model demonstrated strong performance (CCC: 0.93-0.99) for determination of aortic area parameters. Previously, in MESA, intraclass correlation coefficients (ICC) for intra- and inter-observer reproducibility ranged between 0.87-0.99 and 0.56-0.99 for all aortic parameters, respectively (21). Similar deep learning-based analysis models to assess the aorta reported a mean dice coefficient of 94.0%. The deep learning method for aortic analysis presented here has superior performance compared to manual methods and previously published deep learning models. Importantly, application of the deep learning model on a small cohort of TAA external to the MESA set demonstrated close agreement with manually derived segmentation maps as revealed by the high dice coefficients and high correlations with low observed bias and variability. Despite the enlarged area of the ascending aorta in TAA, and with no TAA patients in the training data, the model was able to successful segment the vessel and accurately quantify area in most cases.

Our model demonstrated moderate correlation for CNN-estimated aortic distensibility with that determined from manual segmentations for both the ascending and descending aorta. In particular, model performance was lowest for descending aortic distensibility on the external test set of TAAs. Bland Altman analysis reveals no significant trends in variance or significant positive or negative bias, though there was a larger difference observed for larger distensibility, which is a limitation of our model. There could be several reasons for this observation. Distensibility is a ratio, and as such, accuracy suffers if either the maximum or minimum aortic area is not accurately quantified. In our analysis, the maximum and minimum aortic areas were based purely on aortic geometry and not on the corresponding systolic and diastolic phases of the cardiac cycle. As a result, maximum and minimum areas are taken from different phases of the cardiac cycle for manual and deep learning derived analyses. While not significantly affecting quantification of aortic area, this approach affects distensibility quantification more. The results from our model are in agreement with previous efforts to assess aortic distensibility in MES -- ICC for inter-class variability for aortic distensibility for MESA was 0.56 between two independent expert users. Consequently, compared to two independent users, deep learning derived distensibility was comparable. We suspect that higher resolution imaging and further model training will allow for distensibility to be calculated with higher accuracy.

Cases in which our deep learning model did not provide good segmentations were within expectation. The training data set in the present study utilized only magnitude images, and phase images were not utilized for training. As such, poorly segmented cases in the internal MESA test data set were primarily cases where manual segmentation was performed utilizing both the magnitude and phase images. Incorporation of phase images would likely improve network performance in these cases. Errors in the external TAA test set, however, were primarily isolated to areas of filling defects in the ascending aorta, well described in TAA, and due to abnormal descending aorta shapes, not present in the training set. Of these, the descending aorta segmentations were the primary source of error. We suspect that transfer learning with additional cases of TAA in the training set will likely reduce segmentation errors.

A major strength of the CNN trained in this study is the utilization of training data set derived across multiple sites and multiple vendors. Studies by Bernard et. al have demonstrated that a neural network trained on a heterogeneous training set have increased generalizability (25). Consequently, the site and vendor generalizability of the CNN that result from the MESA training set in this study are important for effective clinical implementation. The model was able to successfully segment the ascending and descending aorta despite slice selection, anatomic variation, and noise on both an internal and an external data set of TAAs. Additionally, time series information was not considered by the network for segmentation. The primary advantage of this approach is expansion of training set size, increasing probability for convergence utilizing traditional deep learning methodologies.

No aortic deep learning segmentation models have been trained on large cohorts of data, analogous to ImageNet, required for both robust model development and generalizability (26, 27). In one study, Bratt et. al reported uniformly successful automated aortic flow determination utilizing U-Net segmentation (26). However, the CNN was trained on 150 patients, who underwent clinical PC-CMR, an appreciably smaller training set compared to the model presented. To our knowledge, this is the first study to have evaluated deep learning segmentation methods for automated aortic analysis on a large, diverse population like MESA. It is our objective that the weights from deep learning model trained in our study may be used as an initialization for future vascular segmentation tasks.

**Limitations.** Our study has several limitations. Despite the large training data set, several limitations result from utilizing the MESA cohort, though many are addressed by inclusion of an external test set of TAAs. Participants in the MESA cohort are mostly normal without incident CVD. As the MESA cohort is relatively elderly (mean age, 64 years at baseline), our model may have limitations in patients with congenital heart disease, which present earlier in life. As suggested, weights from the network trained in the present study can be applied to various clinically relevant cases utilizing transfer learning. Segmentation maps utilizing deep learning fail to account for sub-pixel areas, therefore small geometric differences in segmentation remain. The strength of our study is the large training data set from MESA derived from multiple sites and vendors, allowing for more generalizable weights.

**CONCLUSION**

The present study demonstrates uniformly successful CNN-based ascending and descending aortic segmentation and excellent performance for quantification of aortic area in both an internal MESA test set and an external test set consisting of a cohort of TAA patients. Aortic distensibility from our model demonstrated moderate correlation with that derived from manual segmentations. It is our objective that the weights from deep learning model trained in our study may be used as an initialization for future vascular segmentation tasks. Future studies may involve training and evaluating expansions of the proposed model on images of specific clinical pathologies.

**DECLERATIONS**

**Ethics approval and consent to participate.** All participants gave informed consent for the study protocol, which was approved by the institutional review boards of all MESA field centers and the CMR reading center. MESA field center IRB numbers (WFU - IRB00008492, COL - IRB00002973; JHU – 00001656; UMH - IRB00000438; NWU - IRB00005003; UCLA – 00000172).

**Consent for publication**. All authors have provided consent for publication.

**Availability of data and materials.** The raw data supporting the conclusions of this article will be made available by the authors, without undue reservation, to any qualified researcher.

**Competing Interest.** There is no potential conflict of interest, real or perceived, by the authors. The views expressed in this manuscript are those of the authors and do not necessary represent the view of the National Heart, Lung, and Blood Institute; the National Institutes of Health; or the U.S. Department of Health and Human Services.

**Grant Support.** This research was supported by contracts HHSN268201500003I, N01-HC-95159, N01-HC-95160, N01-HC-95161, N01-HC-95162, N01-HC-95163, N01-HC-95164, N01-HC-95165, N01-HC-95166, N01-HC-95167, N01-HC-95168 and N01-HC-95169 from the National Heart, Lung, and Blood Institute, and by grants UL1-TR-000040, UL1-TR-001079, and UL1-TR-001420 from the National Center for Advancing Translational Sciences (NCATS).

**Authors' contributions.** VJ, JL, and BAV conceived and designed the study. VJ, NK, AR, GTT, KB, and ADC performed measurements and manual segmentations. NK and AR provided the ArtFUN software. VJ and BAV performed statistical analysis. VJ, SK, JL, and BAV wrote the manuscript. SK, CW, and DB provided additional assistance drafting the manuscript. All authors revised the manuscript critically for important intellectual content, and all authors read and approved the final version to be published

**Acknowledgements.** The authors thank the other investigators, the staff, and the participants of the MESA study for their valuable contributions. A full list of participating MESA investigators and institutions can be found at http://www.mesa-nhlbi.org.

15## TABLES

|  | MESA | | | | TAA |
|---|---|---|---|---|---|
| **Characteristics** | **All MESA Participants (n=1872)** | **Train (n=1123)** | **Validation (n=374)** | **Internal Test (n=375)** | **External Test TAA (n=37)** |
| Number of Images | 40418 | 24282 | 8067 | 8069 | 3224 |
| Age, y | 64±10 | 64±11 | 64±10 | 63±10 | 60±15 |
| Men, % | 48 | 47 | 51 | 51 | 84 |
| Height, cm | 166±10 | 165±10 | 166±10 | 166±10 | 175±9 |
| Weight, kg | 76±16 | 76±16 | 76±16 | 78±16 | 79±16 |
| BMI, kg/m$^2$ | 27.6±5.0 | 27.5±5.0 | 27.6±4.9 | 27.8±4.8 | 25.9±4.7 |
| Blood Pressures | | | | | |
| SBP, mmHg | 128±22 | 127±22 | 130±22 | 130±23 | 119±15 |
| DBP, mmHg | 72±10 | 72±10 | 72±10 | 73±11 | 84±10 |
| PP, mmHg | 56±18 | 56±18 | 57±18 | 57±18 | 35±9.2 |
| Heart rate, bpm | 64±10 | 64±10 | 63±10 | 64±10 | 70±15 |
| Aortic Geometric Parameters | | | | | |
| AA Maximum Area (cm$^2$) | 9.0±2.1 | 9.0±2.1 | 8.7±1.9 | 9.0±2.1 | 12.7±6.1 |
| AA Minimum Area (cm$^2$) | 8.3±2.0 | 8.4±2.0 | 8.1±1.8 | 8.4±2.0 | 11.5±5.7 |
| AA Distensibility (x 10$^{-3}$ mmHg$^{-1}$) | 1.5±1.3 | 1.5±1.3 | 1.6±1.2 | 1.5±1.1 | 3.4±1.9 |
| DA Maximum Area (cm$^2$) | 5.3±1.3 | 5.3±1.3 | 5.1±1.3 | 5.4±1.4 | 5.2±2.7 |
| DA Minimum Area (cm$^2$) | 4.8±1.3 | 4.9±1.3 | 4.7±1.2 | 4.9±1.3 | 4.5±2.4 |
| DA Distensibility (x 10$^{-3}$ mmHg$^{-1}$) | 1.8±1.3 | 1.8±1.3 | 1.8± | 2.0±1.5 | 5.3±4.2 |

**Table 1. Baseline Characteristics and Aortic Parameters Stratified by Training, Validation, and Test Groups.** Values are mean (SD) or %. BMI, body mass index; DBP, diastolic blood pressure; PP, pulse pressure; and SBP, systolic blood pressure. AA, Ascending Aorta. DA, Descending Aorta. TAA, thoracic aortic aneurysm

| | Internal Test Set - MESA | | | |
|---|---|---|---|---|
| | **Manual Mean (± SD)** | **DL Mean (± SD)** | **Mean Diff (± SD) (DL - Manual)** | **CCC** |
| **Ascending Aorta** | | | | |
| Maximum Area ($cm^2$) | 9.0±2.1 | 8.7±1.9 | -0.3±0.4 | 0.97 |
| Minimum Area ($cm^2$) | 8.4±2.0 | 8.3±1.9 | -0.1±0.6 | 0.95 |
| Distensibility (x $10^{-3}$ $mmHg^{-1}$) | 1.5±1.1 | 1.5±1.9 | 0.2±10.0 | 0.63 |
| **Descending Aorta** | | | | |
| Maximum Area ($cm^2$) | 5.4±1.4 | 5.2±1.3 | -0.3±0.3 | 0.96 |
| Minimum Area ($cm^2$) | 5.0±1.3 | 5.0±1.3 | -0.03±0.29 | 0.98 |
| Distensibility (x $10^{-3}$ $mmHg^{-1}$) | 2.0±1.5 | 1.8±1.3 | -2.0±10.0 | 0.63 |

**Table 2. Internal Test Set (MESA) Quantitative Aortic Parameters.** Values are mean (SD). Aortic parameters were quantified for two sets of contours for each participant, namely (1) semi-automated contours obtained from ArtFUN with manual adjustments (Manual) and (2) deep learning derived contours with analysis parameters determined from ArtFUN (Deep Learning). CCC was calculated relative to "Manual." CCC - concordance-correlation coefficient. AA – Ascending Aorta. DA – Descending Aorta. DL – Deep Learning.

| External Test Set - TAA | | | | |
|---|---|---|---|---|
| | **Manual Mean (± SD)** | **DL Mean (± SD)** | **Mean Diff (± SD) (DL - Manual)** | **CCC** |
| **Ascending Aorta** | | | | |
| Maximum Area ($cm^2$) | 12.7±6.1 | 12.4±6.0 | -0.3±0.3 | 0.99 |
| Minimum Area ($cm^2$) | 11.5±5.7 | 11.3±5.7 | -0.1±0.6 | 0.99 |
| Distensibility (x $10^{-3}$ $mmHg^{-1}$) | 3.4±1.9 | 4.1±4.0 | 0.4±2.4 | 0.62 |
| **Descending Aorta** | | | | |
| Maximum Area ($cm^2$) | 5.2±2.7 | 4.7±2.3 | -0.6±0.8 | 0.93 |
| Minimum Area ($cm^2$) | 4.5±2.4 | 4.2±2.2 | -0.4±.0.8 | 0.93 |
| Distensibility (x $10^{-3}$ $mmHg^{-1}$) | 5.3±4.2 | 3.9±5.1 | -1.0±5.0 | 0.47 |

**Table 3. External Test Set (TAA) Quantitative Aortic Parameters.** Values are mean (SD). Aortic parameters were quantified for two sets of contours for each participant, namely (1) semi-automated contours obtained from ArtFUN with manual adjustments (Manual) and (2) deep learning derived contours with analysis parameters determined from ArtFUN (Deep Learning). CCC was calculated relative to "Manual." CCC - concordance-correlation coefficient. AA – Ascending Aorta. DA – Descending Aorta. DL – Deep Learning. TAA – Thoracic Aortic Aneurysm.

**FIGURE LEGENDS**

**Figure 1. Train, Validation, and Test Group Split and Study Design.** TAA – Thoracic Aortic Aneurysm

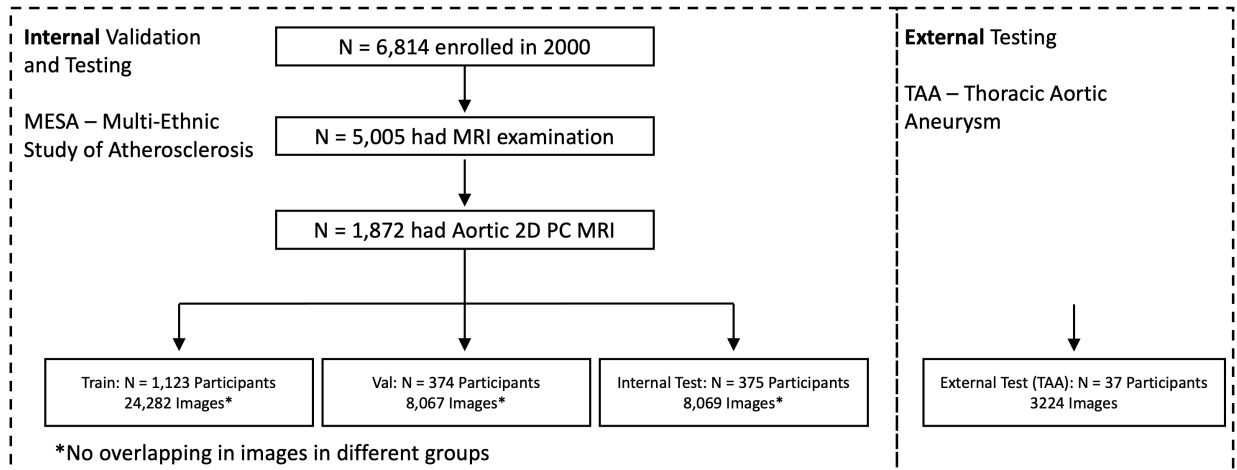

**Figure 2. Dice Coefficient Box Plots for Multi-Class Segmentation Models.** Multi-class segmentation models were trained on (left) non-augmented and (right) augmented training data. Dice coefficients for the ascending and descending aorta were quantified for both multi-class segmentation models on both the internal test set from MESA and external test set from TAAs (Thoracic Aortic Aneurysm). Dice coefficients were evaluated for the entire segmentation as well as individually for the ascending and descending aorta, shown here. AA – Ascending Aorta, DA – Descending Aorta.

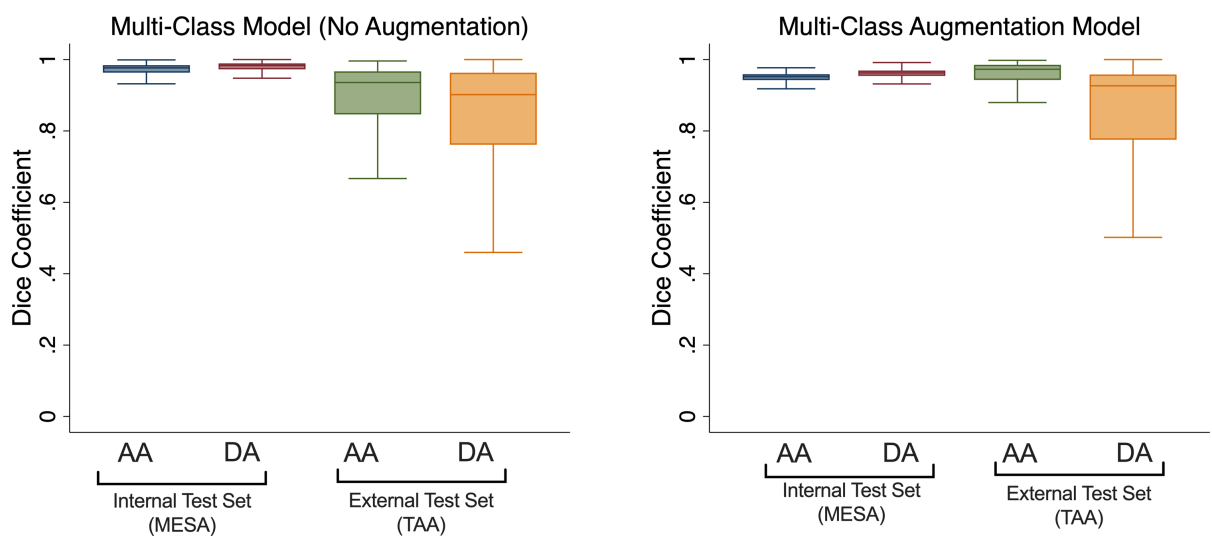

**Figure 3. Representative Deep Learning Aortic Segmentation.** (Upper) Representations of contours after model evaluation on the internal test set from MESA. Green – Ascending Aorta deep learning derived contour. Red – Ascending Aorta manual contour. Orange – Descending Aorta deep learning derived contour. Blue – Descending Aorta manual contour. (Lower) Representations of contours after model evaluation on the external thoracic aortic aneurysm (TAA) test set. Blue – Ascending Aorta deep learning derived contour. Red – Descending Aorta deep learning derived contour.

**Internal Test Set - MESA**

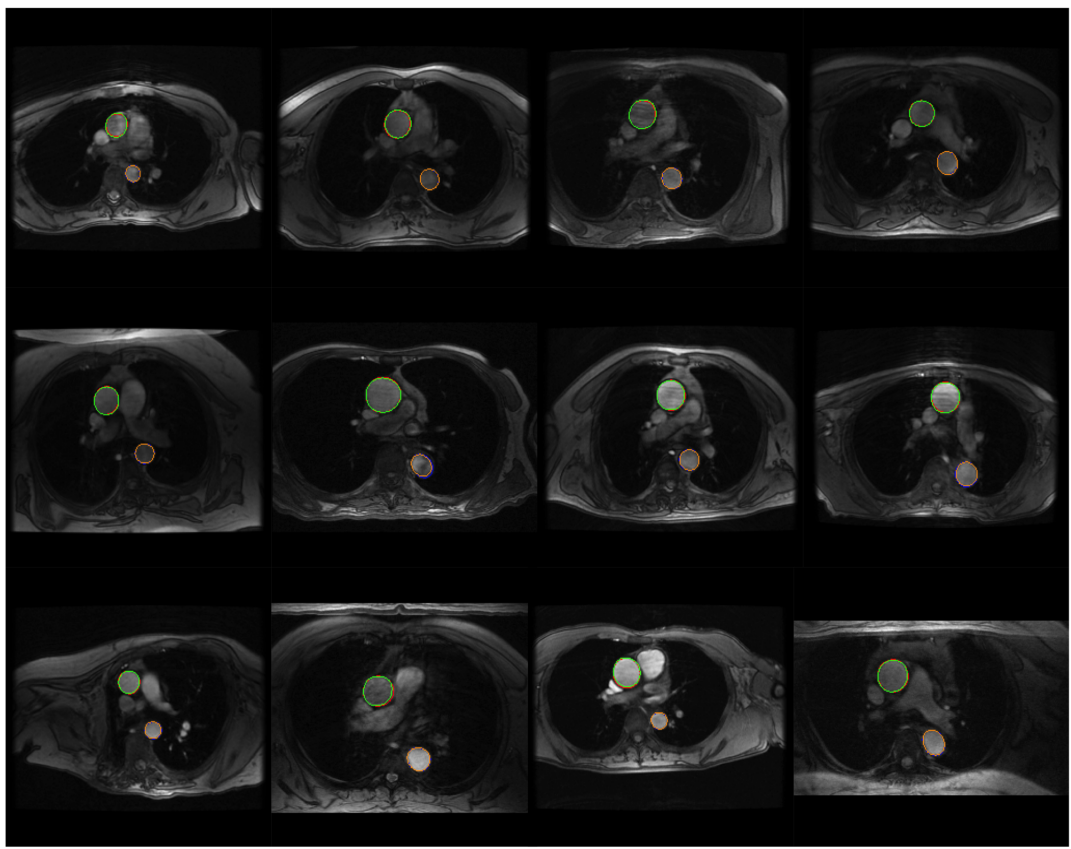

**External Test Set - TAA**

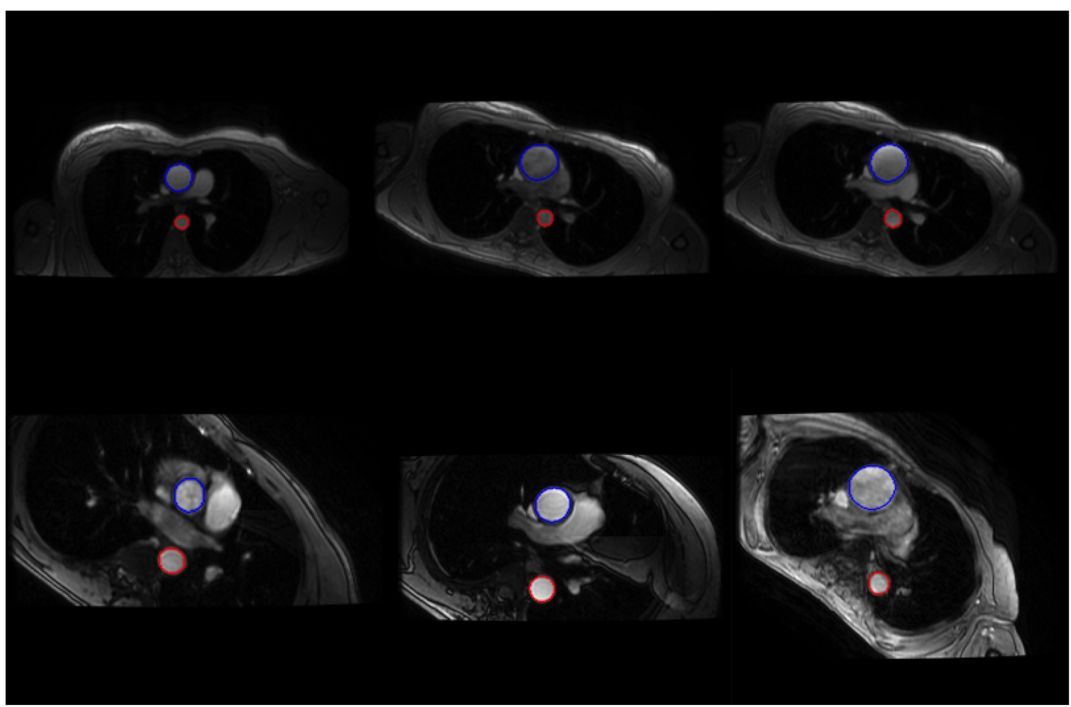

**Figure 4. Bland Altman Analysis Ascending Aortic Quantitative Parameters.** Maximum and minimum ascending aorta area (cm$^2$) along with aortic distensibility (mmHg$^{-1}$) were determined on both the internal MESA derived test set and external thoracic aortic aneurysm (TAA) test set. Shown here are (A) Maximum Ascending Aorta area, (B) Minimum Ascending Aorta area, and (C) Ascending Aortic Distensibility for the internal MESA test set and (D) Maximum Ascending Aorta area, (E) Minimum Ascending Aorta area, and (F) Ascending Aortic Distensibility for the external TAA test set. Corresponding concordance correlation coefficients are found in **Table 2**.

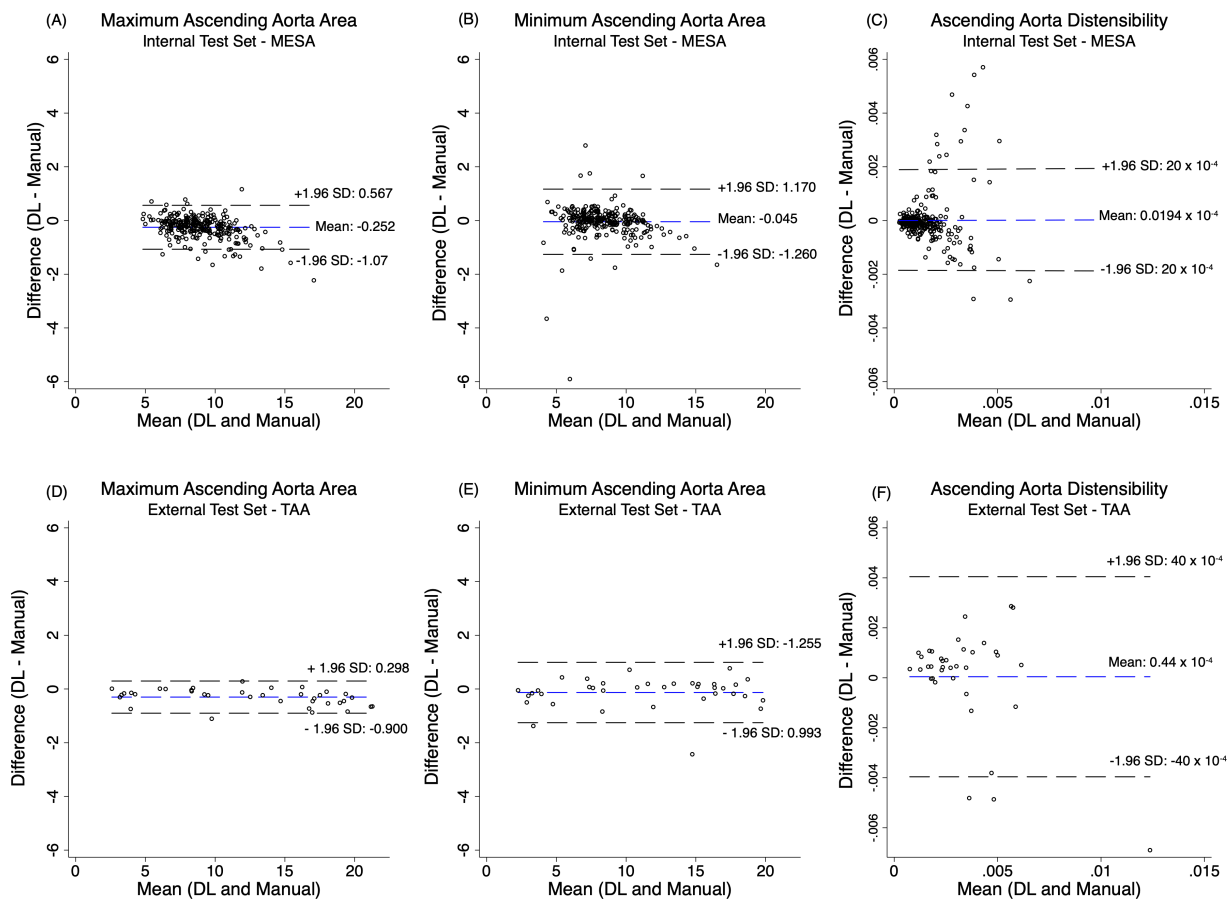

**Figure 5. Bland Altman Analysis Descending Aortic Quantitative Parameters**. Maximum and minimum descending aorta area (cm$^2$) along with aortic distensibility (mmHg$^{-1}$) were determined on both the internal MESA derived test set and external thoracic aortic aneurysm (TAA) test set. Shown here are (A) Maximum Descending Aorta area, (B) Minimum Descending Aorta area, and (C) Descending Aortic Distensibility for the internal MESA test set and (D) Maximum Descending Aorta area, (E) Minimum Descending Aorta area, and (F) Descending Aortic Distensibility for the external TAA test set. Corresponding concordance correlation coefficients are found in **Table 2**.

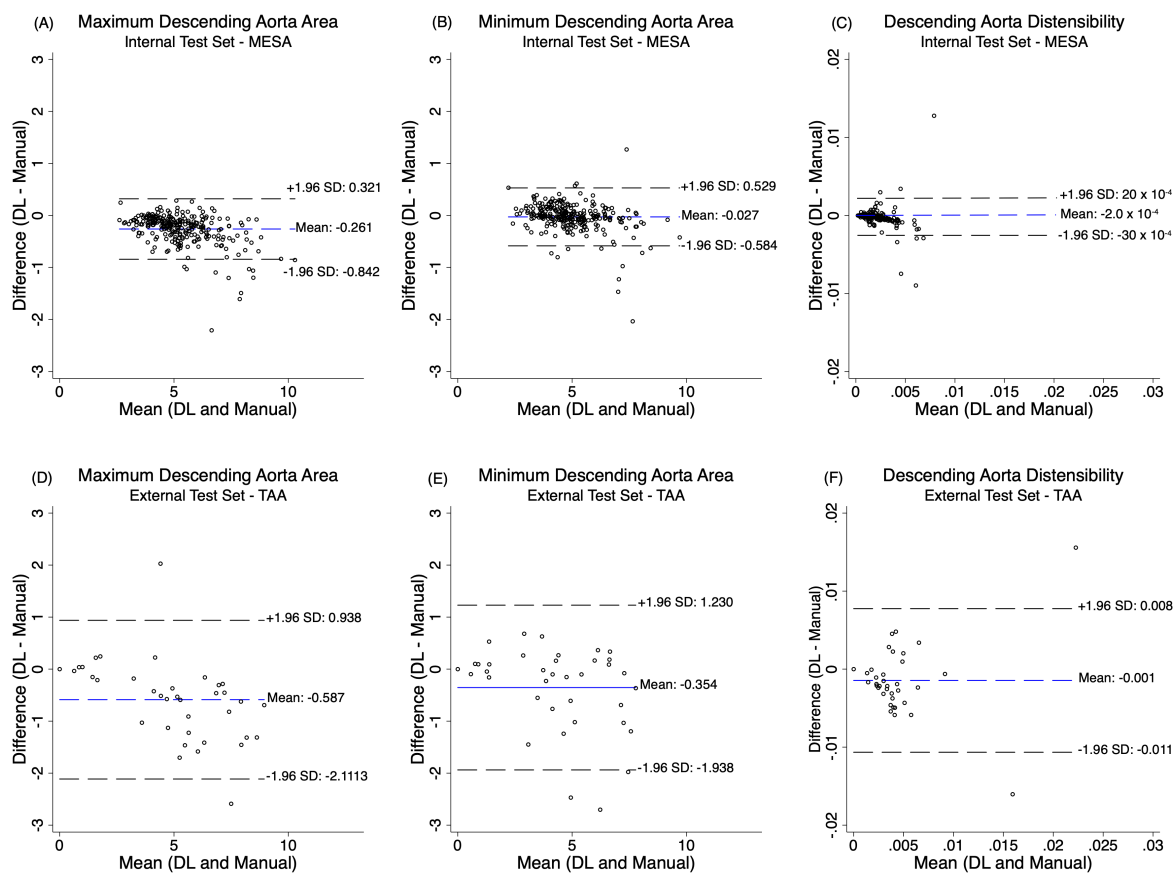

**SUPPLEMENTAL TABLES**

| Model | Average Dice Coefficient | (95% Confidence Interval) |
|---|---|---|
| Binary Segmentation | 97.32% | (97.26%-97.39%) |
| Binary Segmentation w/ Augmentation | 97.60% | (97.52%-97.67%) |
| Multi Class Segmentation | | |
|    Ascending Aorta Only | 96.67% | (96.57%-96.76%) |
|    Descending Aorta Only | 97.64% | (97.59%-97.70%) |
| Multi Class Segmentation w/ Augmentation | | |
|    Ascending Aorta Only | 94.41% | (94.30%-94.53%) |
|    Descending Aorta Only | 95.86% | (95.79%-95.92%) |

**Supplemental Table 1. Internal Model Testing - Dice Coefficients for Binary Segmentation and Multi-Class Segmentation Models.** Both binary class and multi-class segmentation models were trained on non-augmented and augmented training data. Dice coefficients for the ascending and descending aorta were quantified for both multi-class segmentation models. The combined (AA/DA) dice coefficient is the dice coefficient when both the ascending and descending aorta are included in the segmentation simultaneously.

| Site | Participants | Images | AA Dice Coefficient (± SD) | DA Dice Coefficient (± SD) |
|---|---|---|---|---|
| Wake Forest University (WFU) | 47 | 939 | 96.0% ± 6.5% | 97.0% ± 3.0% |
| Columbia University (COL) | 87 | 1740 | 97.0% ± 3.2% | 97.7% ± 3.2% |
| Johns Hopkins University (JHU) | 64 | 1291 | 96.1% ± 4.9% | 97.3% ± 3.3% |
| University of Minnesota, Twin Cities (UMN) | 21 | 488 | 96.9% ± 2.5% | 98.1% ± 2.0% |
| Northwestern University (NWU) | 54 | 1492 | 97.6% ± 1.5% | 98.4% ± 1.3% |
| University of California, Los Angeles (UCLA) | 102 | 2118 | 96.3% ± 5.2% | 97.5% ± 2.4% |

**Supplemental Table 2. Internal Model Testing - Dice Coefficients for Multi-Class Segmentation Model Separated by Site.**